\newcommand{\arcdeg}{\mbox{$^\circ$}}
\newcommand{\halp}{H$\alpha$}

\documentclass{aa}  

\usepackage{graphicx}
\usepackage{txfonts}

\begin{document}


\title{MUSE observations of the counter-rotating \\ nuclear ring in NGC~7742}

\author{Thomas P. K. Martinsson\inst{1,2,3}
   \and Marc Sarzi\inst{4}
   \and Johan H. Knapen\inst{2,3}
   \and Lodovico Coccato\inst{5}
   \and \\ Jes\'us Falc\'on-Barroso\inst{2,3}
   \and Bruce G. Elmegreen\inst{6}
   \and Tim de Zeeuw\inst{5,7,8}
   }
                  
\institute{
  Centro de Astrobiolog\'{i}a (CAB, CSIC-INTA), Ctra de Torrej\'{o}n a Ajalvir, 28850 Torrej\'{o}n de Ardoz, Madrid, Spain\\
  \email{tmartinsson@cab.inta-csic.es}
\and
  Instituto de Astrof\'{i}sica de Canarias, E-38205 La Laguna, Tenerife, Spain
\and
  Departamento de Astrof\'{i}sica, Universidad de La Laguna, E-38206 La Laguna, Tenerife, Spain
\and
  Centre for Astrophysics Research, University of Hertfordshire, Hatfield, Herts AL1 9AB, UK
\and
  European Southern Observatory, Karl-Schwarzschild-Str.\ 2, D-85748 Garching, Germany
\and
  IBM T.\ J.\ Watson Research Center, 1101 Kitchawan Road, Yorktown Heights, NY 10598, USA
\and
  Sterrewacht Leiden, Leiden University, Postbus 9513, NL-2300 RA Leiden, The Netherlands
\and
  Max Planck Institut f\"ur extraterrestrische Physik, Giessenbachstrasse, 85748 Garching, Germany  }


%
\abstract{}
%
{We present results from MUSE observations of the nearly face-on disk galaxy NGC~7742.
This galaxy hosts a spectacular nuclear ring of enhanced star formation, which is unusual
in that it is hosted by a non-barred galaxy, and also because this star formation is most
likely fuelled by externally accreted gas that counter-rotates with respect to its main
stellar body.}
{We use the MUSE data to derive the star-formation history (SFH) and accurately measure
the stellar and ionized-gas kinematics of NGC~7742 in its nuclear, bulge, ring, and disk
regions.}
{We map the previously known gas counter-rotation well outside the ring region and deduce
the presence of a slightly warped inner disk, which is inclined $\sim$6 degrees compared
to the outer disk.
The gas-disk inclination is well constrained from the kinematics; the derived inclination
13.7 $\pm$ 0.4 degrees agrees well with that derived from photometry and from what one
expects using the inverse Tully-Fisher relation.
We find a prolonged SFH in the ring with stellar populations as old as 2--3~Gyr and an
indication that the star formation triggered by the minor merger event was delayed in the
disk compared to the ring.
There are two separate stellar components: an old population that counter-rotates with the
gas, and a young one, concentrated to the ring, that co-rotates with the gas.
We recover the kinematics of the old stars from a two-component fit, and show that
combining the old and young stellar populations results in the erroneous average
velocity of nearly zero found from a one-component fit.}
{The superior spatial resolution and large field of view of MUSE allow us to establish the
kinematics and SFH of the nuclear ring in NGC 7742. We show further evidence that this
ring has its origin in a minor merger event, possibly 2--3 Gyr ago.}
%


 \titlerunning{MUSE observations of NGC 7742}
 \authorrunning{Thomas P. K. Martinsson et al.}
 \maketitle

%
\section{Introduction}
Nuclear rings are important structures for the evolution of disk galaxies. The rings are
prime tracers of disk galaxy dynamics and give information on the underlying dynamical
structure of galaxies. They are also indicators of angular momentum transport, a key
factor in secular evolution.

The advantages of studying rings are that they are common, bright, and easily observable.
Star-forming nuclear rings occur in $\sim$20\% of disk galaxies and are found in S0--Sd
galaxies with a peak in Sab--Sb galaxies \citep{knapen2005,comeron2010}.
They are the site of intense star formation, with rates $\sim$100 times what is observed
in galactic disks \cite[e.g.,][]{hsieh2012}.

Nuclear rings are generally thought to form as gas accumulates at special orbital
resonances while it is channeled toward the central regions of a galaxy by bars
\citep{combes1985,heller1994,knapen1995}.
Their formation, evolution, and demise are therefore central to the secular growth of the
central stellar-mass concentrations of galaxies and may be responsible for the creation of
so-called pseudo-bulges \citep{kormendy2004}.
They are also relevant to the fueling of central supermassive black holes, as they
essentially halt such fueling, and more generally to the connection between AGN and
central star-formation activity \citep{kauffmann2003b}.

%
\begin{figure*}[t]
\centering
\includegraphics[width=0.96\textwidth]{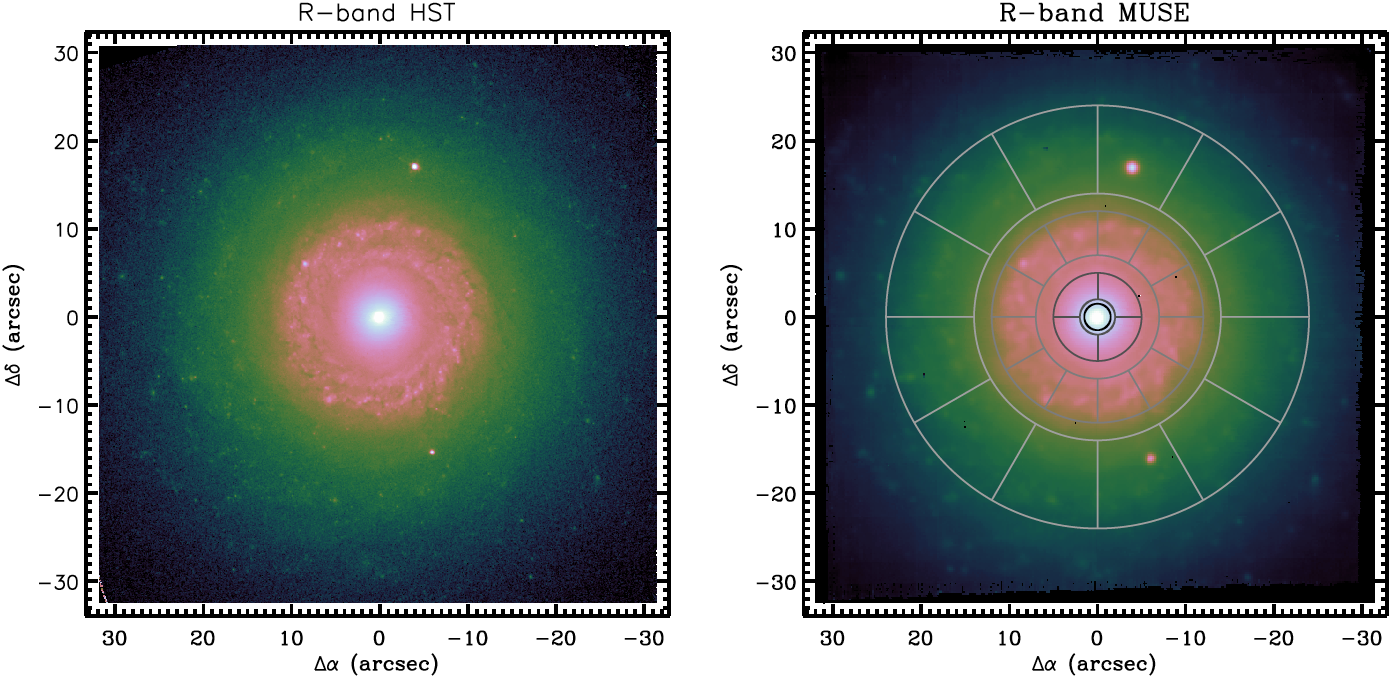}
\caption{HST versus MUSE. 
{\bf Left:} Hubble Space Telescope R-band image of NGC~7742. North is up and east is to
the left. The 60\arcsec\ side is $\sim$7~kpc wide in physical scale, assuming a distance
to NGC~7742 of 22~Mpc.
{\bf Right:} MUSE recontructed R-band image of NGC~7742, with the locations of 29
apertures overplotted. These apertures were used to extract the high-S/N spectra that
provided the optimal stellar templates at the basis of our stellar kinematic and gas
measurements. The apertures were designed to sample the nuclear, bulge, ring, and disk
regions, allowing us to also investigate the SFH across these regions.
The two foreground stars were masked out during the extraction of the aperture spectra.
}
\label{fig:psf_Apertures}
\end{figure*}

Of order 20\% of the star-forming nuclear rings occur in disk galaxies where a stellar
bar is absent \citep{comeron2010}. NGC~7742 belongs to this minority group.
NGC~7742 is a nearly face-on SA(r)b galaxy with a morphology dominated by its
nuclear ring. The ring has a radius of 9\arcsec\ \citep{comeron2010}, or $\sim$1.0~kpc
assuming the distance to the galaxy $D=22$~Mpc \citep{mazzuca2006}. The size of the
galaxy disk is $R_{25}=52$\arcsec\ (or $\sim$6.0~kpc).
There is no evidence for current interactions with other galaxies, although it may form a
pair with NGC~7743, which is 320~kpc away and within 50~km~s$^{-1}$ in radial velocity
\citep{mazzuca2006}.

\cite{buta1996} stated that most non-barred nuclear-ring galaxies show evidence for an
alternative mechanism to obtain non-axisymmetry in the  gravitational potential, such as
strong spiral patterns, weak oval distortions, or disturbances due to a companion galaxy.
Minor merger events have also been suggested to trigger the formation of nuclear rings in
unbarred galaxies \citep[e.g.,][]{knapen2004}, and NGC~7742 is an ideal case for testing
such ideas given that a counter-rotation between the gas in the ring and the stars in the
bulge \citep[][]{deZeeuw2002} reveal an external origin for this nuclear ring.
The ring was interpreted as a result of a minor merger already by \cite{mazzuca2006} and
\cite{Silchenko_Moiseev2006}. That interpretation seems adequate, since a major merger
would completely destroy the disk while anything less than a minor merger would have
difficulties resulting in counter-rotation in the central region. 

Investigation of galaxies with counter-rotating components, such as NGC~7742, is
important for understanding galaxy evolution, as it gives us information on previous
merger and accretion events and the role of these events in the evolution of galaxies.
For example, mergers have been shown as a mechanism for the formation of S0 galaxies
\citep[e.g.,][]{bekki1998},
which could explain why counter-rotation is more common in these morphological types
\citep[][]{CorsiniBertola1998, corsini2014, bassett2017}.
NGC~7742 is again a special galaxy here, in that only about 10\% of spiral galaxies have
counter-rotating disks \citep{pizzella2004}.

The early SAURON integral-field maps of \cite{deZeeuw2002} first presented evidence for an
external origin of the nuclear ring of NGC~7742. Later SAURON measurements of
\cite{Falcon-Barroso2006} served to trace the transition in the gas properties from the
nucleus into the ring when they were combined with DensePak data in \cite{mazzuca2006}.
\cite{sarzi2007} placed constraints on the star-formation history of the nuclear ring
of NGC~7742 using long-slit measurements and, like for other objects in their study,
found that this ring formed most likely over a prolonged period of time (a few 100 Myr)
characterized by episodic bursts of star-formation activity
\citep[see also][]{allard2006}.

\cite{mazzuca2006} used SAURON and DensePak data and only covered the central
1.5~kpc of NGC 7742.  Thanks to the relatively large field of view (FoV) of MUSE, we go
out further in the disk, reaching $R > 3$~kpc. Together with the high spatial resolution
and extended wavelength range, the MUSE data allow us to separate the stellar-population
properties of the bulge, ring, and disk regions and to robustly separate the kinematics of
the main stellar body of NGC~7742 from that of the newly-acquired gas and the stars that
formed from it.

After a short description of the observing strategy, data reduction, and data analysis in
Sect.~\ref{sec:ObsDatRed}, we present our results in Sect.~\ref{sec:Results}.
In Sect.~\ref{sec:Conclusions}, we summarize these results.
%

%
\section{Observations, reduction, and analysis}
\label{sec:ObsDatRed}

%
\begin{figure*}
\centering
\includegraphics[width=0.96\textwidth]{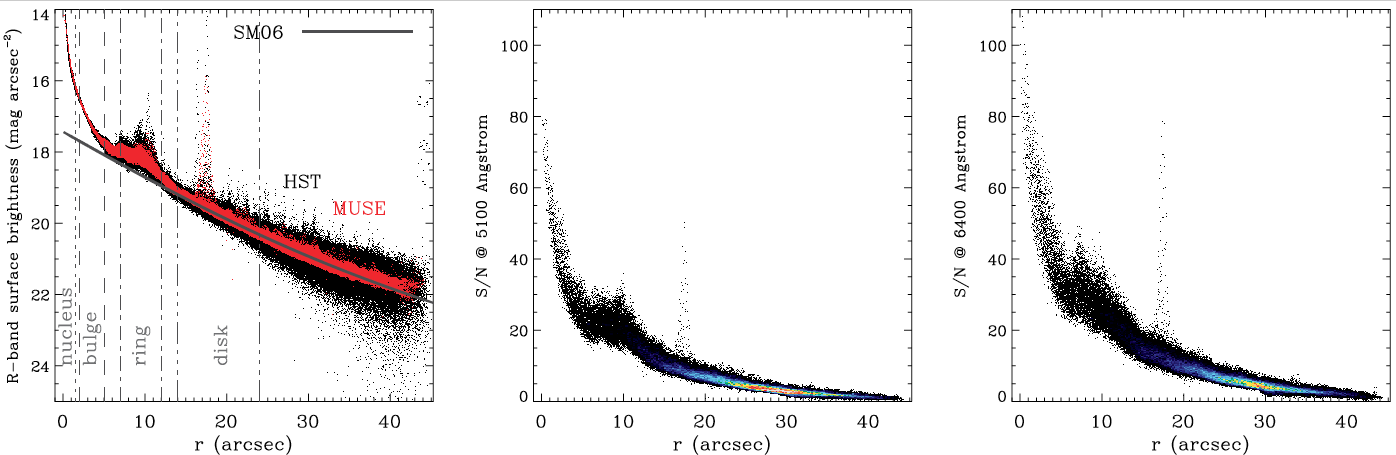}
\caption{Surface brightness and S/N in individual spaxels as a function of radius.
{\bf Left:} R-band surface brightness as a function of radius, from HST (black dots)
and MUSE (red dots) data. The grey line indicates the two-disk model of
\cite{Silchenko_Moiseev2006} rescaled to R-band, highlighting the regions dominated by the
disk surface brightness. The radial extent of the apertures shown in
Fig.\ref{fig:psf_Apertures} are also indicated.
{\bf Middle:} S/N as a function of radius and per spectral resolution element,
at 5100~\AA.
{\bf Right:} same as middle, but now at 6400~\AA.
The colours in the middle and right panels indicate the density of individual spaxel
measurements (redder colours means higher density).
}
\label{Fig:SNa}
\end{figure*}

%
\subsection{Observing strategy}
Observations of NGC~7742 were carried out with the Multi-Unit Spectroscopic Explorer
\citep[MUSE;][]{bacon2010} on June 22, 2014, during the instrument science verification
in the context of Programme 60.A-9301 (PI Sarzi).
MUSE is a second generation integral field spectrograph that consists of 24 identical
integral field units (IFUs), and operates on the Unit Telescope 4 (UT4) of the Very Large
Telescope (VLT) in Paranal, Chile.

We used MUSE in the wide-field spatial mode and the nominal wavelength mode. The
wide-field mode covers a $60$\arcsec$\times60$\arcsec\ FoV with a spatial sampling of
$0\farcs2 \times 0\farcs2$. The spectroscopic nominal mode covers a 4800\AA\ -- 9300\AA\ 
wavelength range with a dispersion of 1.25\AA\ pixel$^{-1}$ and a nominal resolving power
that ranges from 1770 (at 4800\AA) to 3590 (at 9300\AA).

The observations were organized in two exposures of 1800s, offset by 1\arcsec\ in
declination and rotated by 90\arcdeg. One offset exposure of 300s was taken in between
for the evaluation of the sky background.
As part of the standard MUSE calibration plan, a set of bias, internal flat fields,
sky flat fields, and arc-lamp exposures were taken, and a spectro-photometric standard
star was observed for photometric calibration.
%

%
\subsection{Data reduction}
We used the MUSE pipeline \citep[v1.6;][]{weilbacher2015} within the Reflex environment
\citep{freudling2013} to reduce the data.
For each on-sky exposure, we used the flat fields that were taken within 12 hours from the
scientific exposure, and that were closer in temperature. In our case, flats closer in
temperature to the observations were also closer in time.
Large-scale illumination variations were corrected via sky flat fields, obtained during
twilight.
The white dwarf star GD~153 was observed at twilight with the same observational set-up of
NGC~7742, and it was used to correct for the instrument response and telluric absorption.

The contributions of the night sky emission lines and continuum were evaluated from the
dedicated offset exposure. The pipeline constructed a model of the sky spectrum and
subtracted it from the two scientific exposures, after scaling it for exposure time and
after convolving it for the instrumental line-spread function for each slice and IFU.

Figure~\ref{fig:psf_Apertures} (left panel) shows the HST R-band image, which can be
compared to the reconstructed MUSE R-band image (right panel).
The comparison illustrates the quality of the data. As can be seen, the seeing was very
good (FWHM $\sim 0\farcs6$, measured from the two foreground stars) during the observations.

Figure~\ref{Fig:SNa} further illustrates the quality of the MUSE data by comparing
the surface brightness of the HST and the reconstructed MUSE images as well as showing the
run of signal-to-noise ratio (S/N) in each individual MUSE spectrum (that is prior to any
spatial binning) as a function of radius.
%

%
\subsection{Data analysis}
\label{sec:Analysis}
In order to extract the stellar and gas kinematics and measure the gas flux distribution
in NGC~7742, we first proceeded to spatially bin our MUSE cube using the 
Voronoi tessellation method of \cite{cappellari2003} and a formal target S/N$=$30 in each
bin.
Such a S/N target is sufficient to accurately measure the stellar kinematics over the
relatively long wavelength range afforded by the MUSE spectra\footnote{We settled on
a S/N=30 target for our spatial binning after measuring the stellar kinematics on the
MUSE data and observing how the formal pPXF errors on the velocity and velocity dispersion
fare as a function of S/N. This experiment showed that, within our chosen wavelength range
and with an appropriate set of pre-defined optimal templates (such as we obtain from our
apertures), by a S/N=30 we reach a precision of $\sim$10~km/s and $\sim$15~km/s in the
stellar velocity and velocity dispersion, respectively. This is also broadly consistent
with that found from simulations of SAMI data by \cite{fogarty2015}, which is not
surprising given that they also have a fairly long wavelength range that includes several
stellar absorption features.}, which in turn ensures robust emission-line measurements.
For this purpose, we used the penalized pixel fitting (pPXF) code of
\cite{cappellari2004} to extract the stellar kinematics and subsequently measure the
emission-line flux and kinematics using the GandALF code of \cite{sarzi2006}.

When running the pPXF and GandALF procedures on the Voronoi-binned spectra we used a
predefined set of optimal stellar templates that were based on the full MILES
single-age stellar-population models of \cite{Falcon-Barroso2011}. More specifically,
these optimal templates correspond to the best combination of the MILES models that were
needed during the pPXF and GandALF fit to 29 high-S/N aperture spectra covering the
central 24\arcsec\ of NGC~7742, where we obtained an optimal template for each of
the 29 individual apertures.
As shown in Fig.~\ref{fig:psf_Apertures}, these apertures were designed to sample the
nuclear, bulge, ring, and disk regions of NGC~7742, where variations of the
stellar-population properties are more likely to occur. In fact, the analysis of such
aperture spectra also allowed us to estimate the star-formation history in these four
regions (Sect.~\ref{sec:StarPops}).

It should be noted that the naming of the different regions is only indicative. The
``ring'' apertures were based on the extent of the ring in the \halp\ image. The ``disk''
apertures were extracted from the same number of apertures out to a radius where we only
include spectra with sufficient S/N. The ``bulge'' region is taken to be inside the ring
and subdivided into four regions, plus a central ``nuclear'' region.
However, the two-disk model of \cite{Silchenko_Moiseev2006} in Fig.~\ref{Fig:SNa} shows
that what we call ``disk'' apertures are indeed dominated by the disk light, whereas the
regions inside the ring see a substantial contribution from the bulge component.
%

%
\begin{figure*}
\centering
\includegraphics[width=1.0\textwidth]{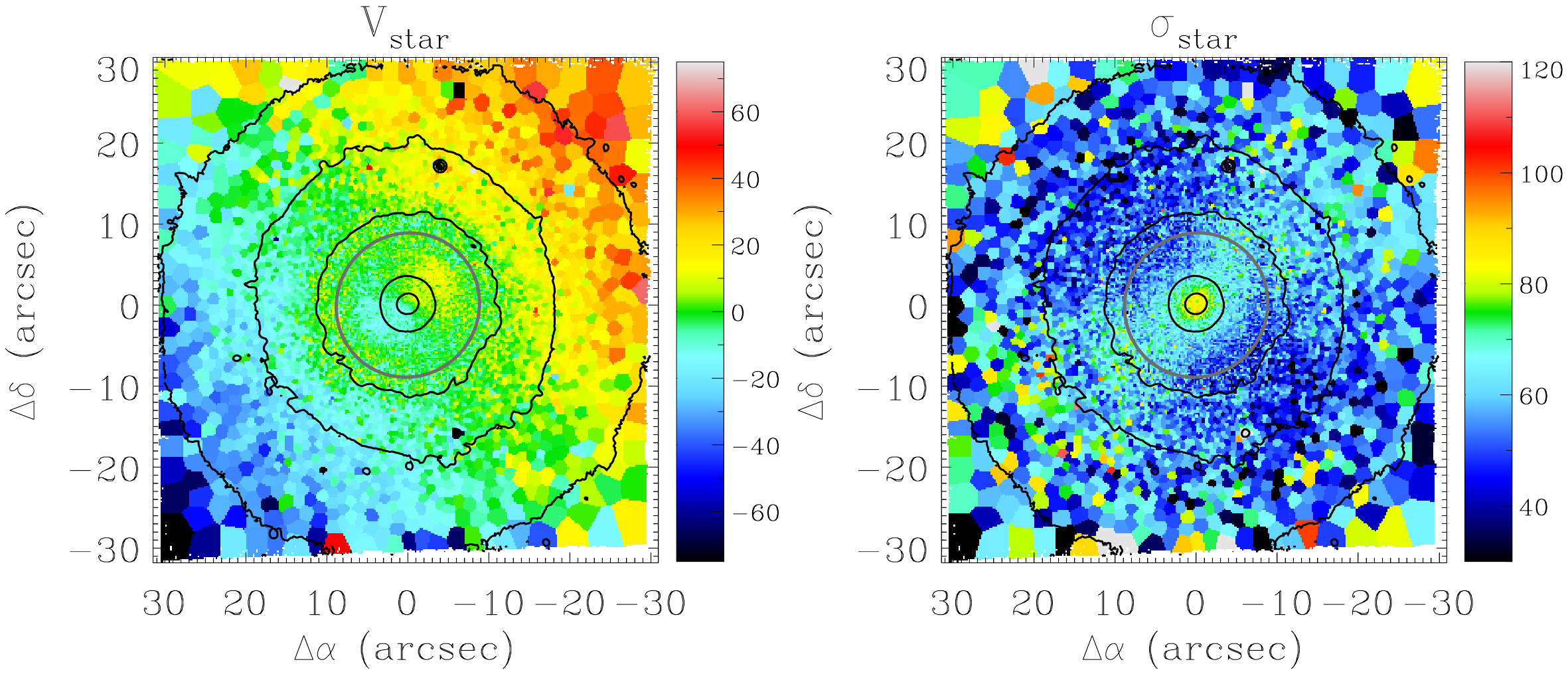}
\caption{Stellar kinematics.
{\bf Left:} Stellar velocity field.
{\bf Right:} Stellar velocity dispersion map.
The black contours indicate surface brightness levels in the Hubble F675W image
(Fig.~\ref{fig:psf_Apertures}),
in intervals of half a magnitude. The grey ellipse in each figure at
R$=$9\arcsec\ indicates the location of the ring as defined by \cite{comeron2010}.
}
\label{fig:star_kin}
\end{figure*}

%
\section{Results}
\label{sec:Results}
%
\subsection{Stellar kinematics}
\label{sec:StarKin}
Figure~\ref{fig:star_kin} shows the stellar velocity field and the stellar velocity
dispersion map as obtained from our pPXF fits. These maps provide a clearer and more 
complete view of the stellar kinematics of NGC~7742 as first mapped by 
\cite{Falcon-Barroso2006}, and also clearly show how the previously reported
counter-rotation between stars and gas extends well beyond the ring region
(see gas maps in Fig.~\ref{fig:gas_kinematics}).

The MUSE data also bring out much more clearly the kinematic presence of the newly-formed
stars in the ring, and possibly also in the disk regions. As these new stars rotate in an
opposite direction to that of the bulge stars, our pPXF measurements for the relative
velocity estimates decrease to near-zero values along the kinematic major axis,
where we also observe a similarly artificial increase in the stellar velocity dispersion.
The pPXF measurements shown in Figure~\ref{fig:star_kin} are indeed based on the
assumption that a single velocity distribution would apply to all stars observed along the
light-of-sight, which would then lead pPXF to estimate an average, nearly zero velocity in
the presence of two counter-rotating stellar components, and to be similarly biased
towards larger values for the velocity dispersion \citep{bertola1996, krajnovic2011}.
In order to show how the kinematics of these components can be separated, in
Sect.~\ref{sec:RingDec} we will use priors on the stellar-population properties for the
main and newly-born stellar population of NGC~7742.

We confirm a rather low stellar velocity dispersion in the centre
($\sigma_{*,0} <$~90~km/s) as reported by \cite{Silchenko_Moiseev2006}; for being of an
early type, this galaxy is of rather low mass with only a minor inner mass concentration.
The photometric disk decomposition in \cite{Silchenko_Moiseev2006} also results in
only a small and modest bulge component.
%

%
\subsection{Gas kinematics}
\label{sec:GasKin}
\begin{figure*}[t]
\centering
\includegraphics[width=1.0\textwidth, angle=0]{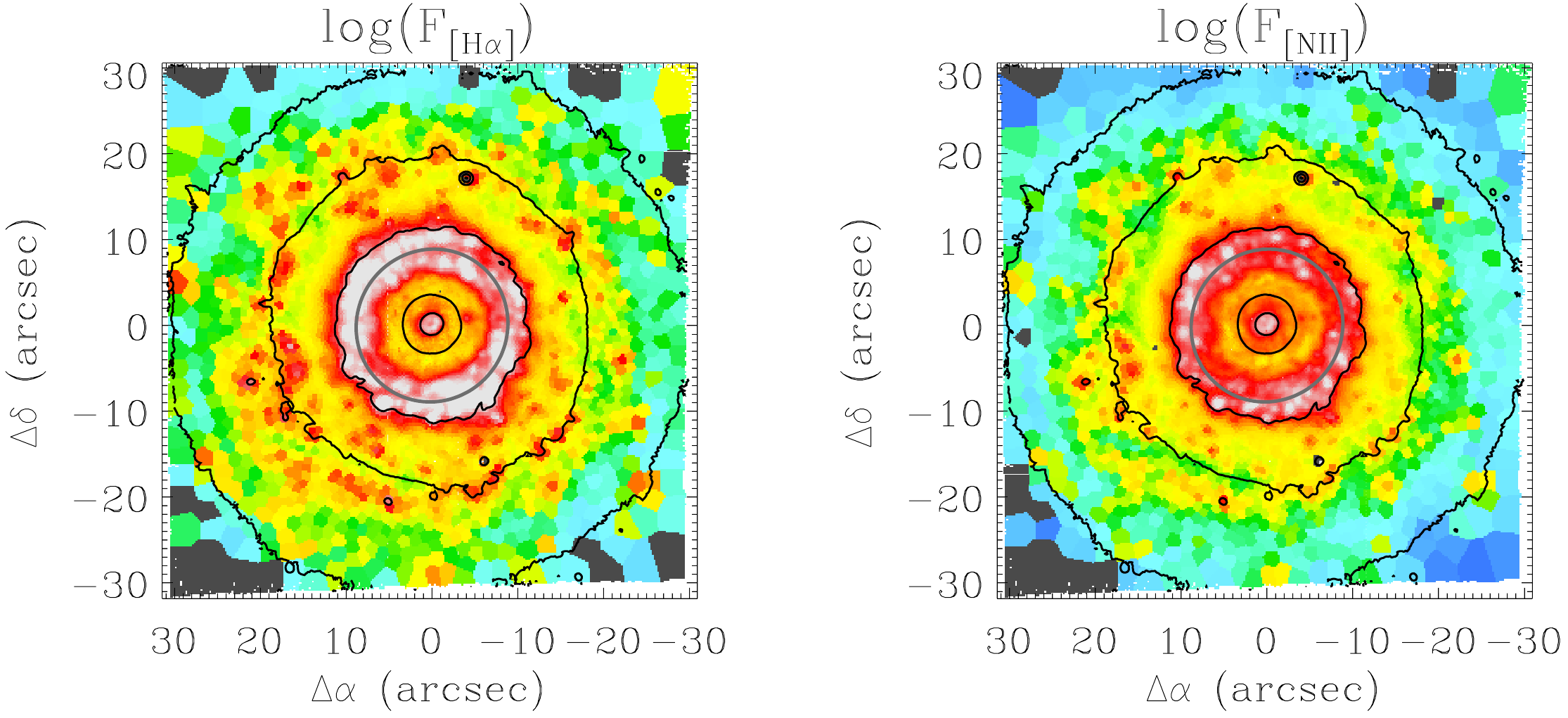}
\caption{Ionized-gas emission flux.
{\bf Left:} \halp\ flux map.
{\bf Right:} [NII] flux map.
Redder colours indicate a higher flux, with white (seen in the ring region)
representing the highest flux.
Contours are the same as in Fig.~\ref{fig:star_kin}.
Note the stellar cluster inside the ring, 4\arcsec\ to the west (right) of the
centre, most clearly seen in the \halp\ map, and also seen as a dip in the velocity
dispersion in Fig.~\ref{fig:gas_kinematics}.
}
\label{fig:gas_flux}
\end{figure*}

In Fig.~\ref{fig:gas_flux}, we plot the ionized-gas ([NII] and \halp) flux maps. These
maps show that there is plenty of gas in this galaxy, with a concentration in the
star-forming ring. The \halp\ flux, in particular, is clearly stronger where star
formation is taking place.
From these maps, more than 100 stellar clusters can be detected, with a particularly
interesting one being located inside the ring, just 4\arcsec\ west of the centre
seen most clearly in the \halp\ map.
This stellar cluster proves that the star formation is ongoing also well inside the
ring.

\begin{figure*}
\centering
\includegraphics[width=1.0\textwidth]{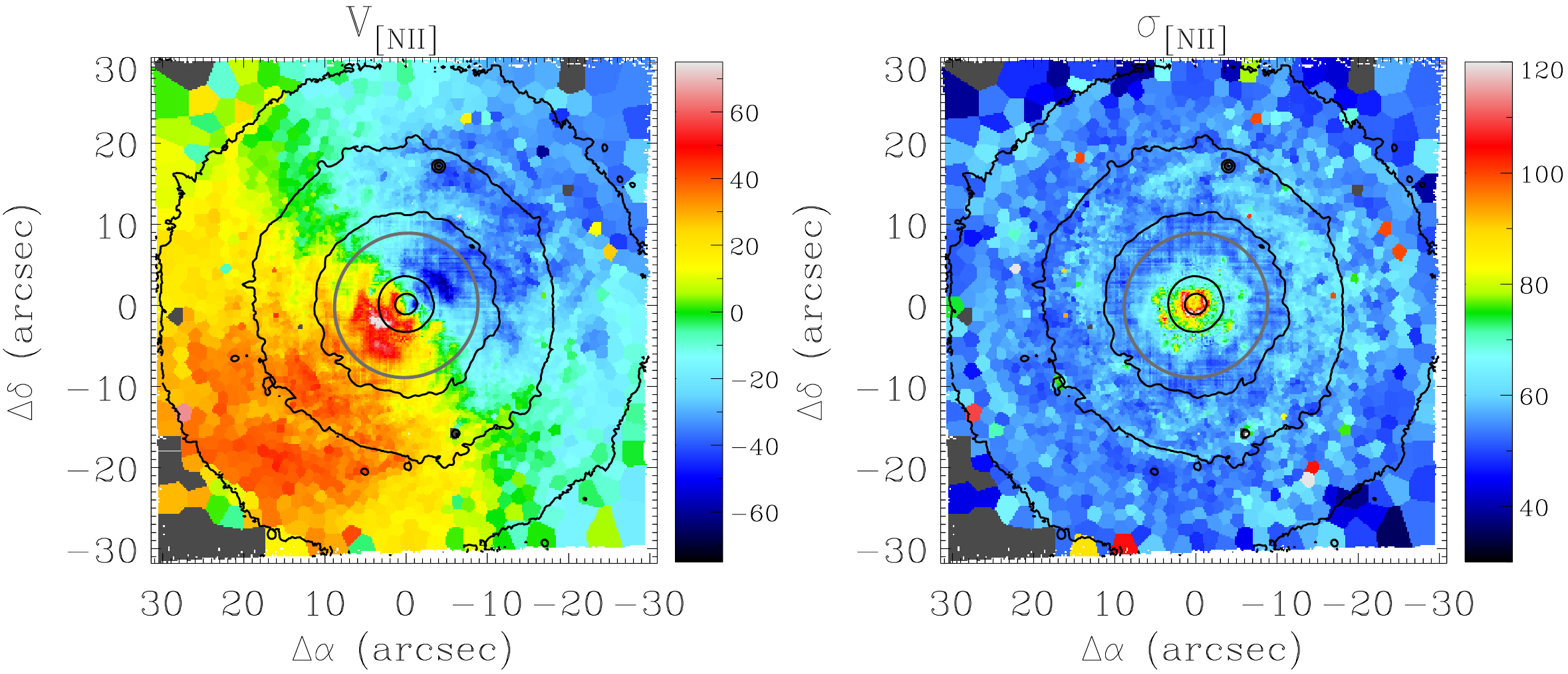}
\caption{Ionized-gas kinematics.
{\bf Left:} Velocity field traced by the [NII] forbidden lines.
{\bf Right:} [NII] gas velocity dispersion.
Contours are the same as in Fig.~\ref{fig:star_kin}.
}
\label{fig:gas_kinematics}
\end{figure*}

Figure~\ref{fig:gas_kinematics} shows the ionized-gas velocity field and velocity
dispersion map (produced for both [NII], where all forbidden lines were tied
together, and \halp, which show very similar kinematics; only the [NII] velocity field and velocity dispersion maps have been represented in Fig.~\ref{fig:gas_kinematics}).
The velocity field displays rather regular rotation over the full FoV, although one can
note that
it sometimes deviates significantly from that of a disk of gas in circular and
coplanar orbits.
There is a peak in the rotational velocity at about 6\arcsec\ from the center,
but as we will see in Sect.~\ref{sec:warp} this is likely a projection effect from
a warped inner disk.

The gas velocity dispersion map shows a dip in the dispersion at the location of the ring,
which is likely due to the higher star-formation activity and the associated requirement
for gas to be in a dynamically cold state.  Interestingly, a lower gas velocity dispersion
can also be seen in the bulge stellar cluster mentioned above.

In the very centre, inside 2\arcsec, there is a possible decoupled disk with a different
position angle than the outer disk. This seems also associated with an inner ``ring'' of
higher gas dispersion (located approximately at the most inner contour).
However, note that this may also be an AGN-related feature.

%
%
%
%
%

%
\subsection{A possible inner warp}
\label{sec:warp}
On the major axis of the gas velocity fields (Fig.~\ref{fig:gas_kinematics}), we note that
the line-of-sight velocity is higher at about $R=6$\arcsec, then decreases when going
out to larger radii before becoming nearly constant.
A rapid rise of the rotation curve, followed by a dip before reaching its flat part, is
often seen in massive disk galaxies with a major inner mass concentration
\citep[e.g.,][]{sofue1999}.
However, for NGC~7742, which has a modest rotational velocity of $\sim$150~km s$^{-1}$
(see below) and does not appear to have a massive bulge (Sect.~\ref{sec:StarKin}), we do
not expect such a large amplitude difference.
A possible explanation could  be that we see a projection effect: a small warp in
this nearly-face-on galaxy will give large relative differences in the derived velocity.

To investigate this further, we perform a tilted-ring analysis using {\tt ROTCUR} within
the Groningen Image Processing SYstem (GIPSY) software package
\citep{hulst1992,vogelaar2001}.
In principle, all ring parameters can be fit simultaneously. In practice, however, there
are degeneracies affecting the fit, in particular a significant covariance between the
inclination ($i$) of the ring and its rotational velocity ($V_{\rm rot}$) for galaxies
seen nearly face-on. This degeneracy may yield large inclination errors for galaxies with
$i\lesssim40$\arcdeg\ \citep{begeman1989}.
However, \cite{andersen2013} showed that high-quality optical IFU kinematic data can yield
accurate and precise kinematic inclinations for a non-warped disk down to about 15\arcdeg\
when modeling the entire velocity field as a single, inclined disk.

Here we use traditional ring-fitting, fitting the various parameters [centre, systemic
velocity ($V_{\rm sys}$), position angle (PA), $i$, and $V_{\rm rot}$] in five steps,
similar to the method described in \cite{martinsson2013a}.
Due to the degeneracy mentioned above, we did not expect to derive a useful kinematic disk
inclination, but instead rely on the inclination calculated from photometry or inferred
from the inverse Tully-Fisher relation, which is generally more precise in this
inclination regime. However, after having derived $V_{\rm sys}$ and PA\footnote{We
find a gas-disk PA=140$\pm$4 degrees for both tracers, excluding the inner measurements
($R<8$\arcsec).}, the inclination measurements in the rings give consistent results, with relatively small formal errors.

\begin{figure}
\centering
\includegraphics[width=0.50\textwidth]{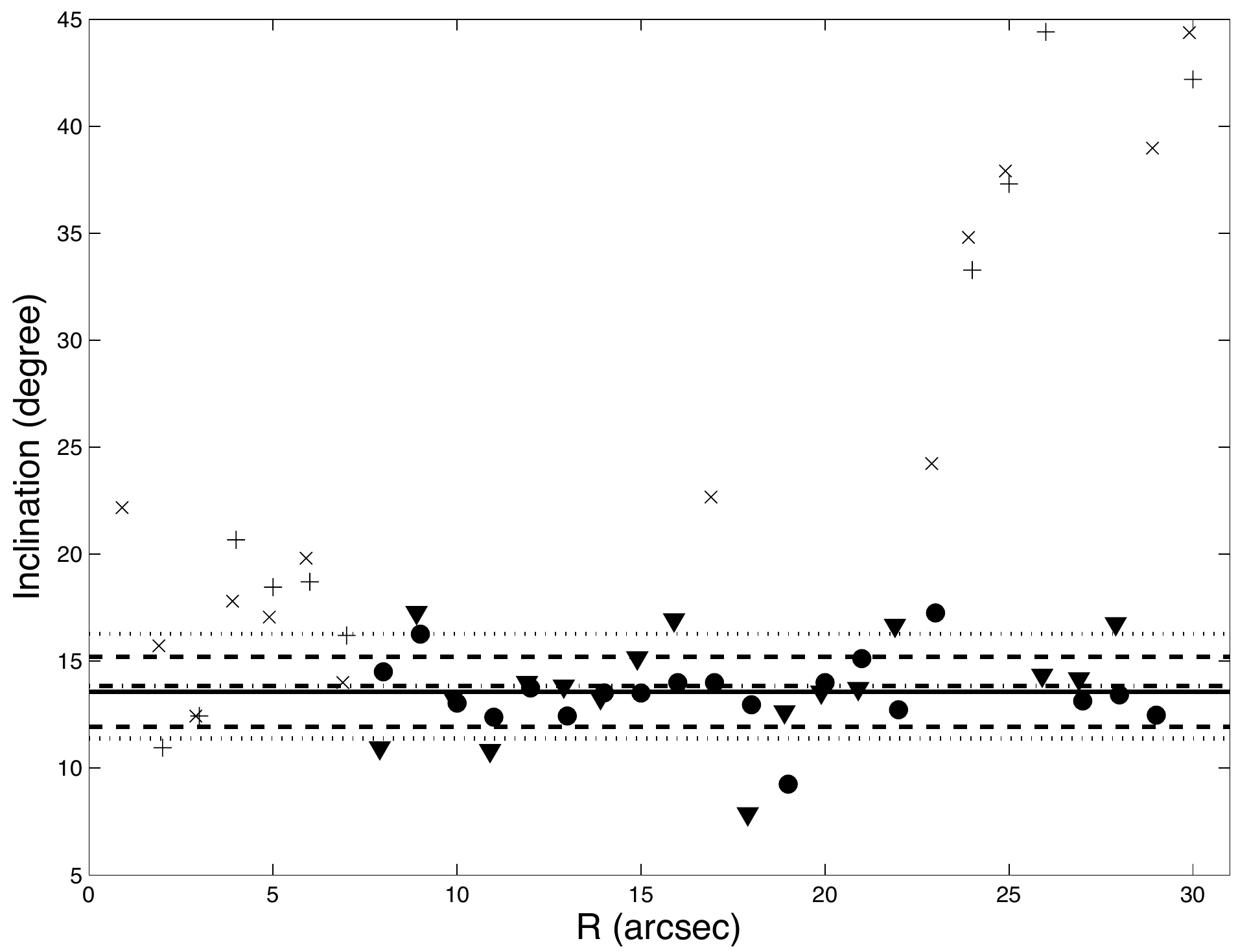}
\caption{Derived disk inclination of the [NII] and \halp\ gas disks.
Circles and triangles show derived inclinations in 1\arcsec\ wide rings for the [NII] and
\halp\ gas, respectively.
Measurements that have been excluded are indicated with crosses ([NII]) and pluses (\halp).
The solid and dashed lines show the calculated mean and standard deviation, respectively,
for the [NII] gas.
The dashed-dotted and dotted lines show the calculated mean and standard deviation,
respectively, for the \halp\ gas.
}
\label{fig:i_kin}
\end{figure}

Figure~\ref{fig:i_kin} shows the disk inclination of the [NII] and \halp\ gas disks
derived from our {\tt ROTCUR} fits in 1\arcsec\ wide rings.
With the exception of the  inner disk and some outliers in the outer disk, we find
consistent inclination values in the rings for both [NII] and \halp. For the [NII] and
\halp\ disks we derive an average inclination and formal error in the mean of
13.6$\pm$0.4 degrees and 13.8$\pm$0.6 degrees, respectively, with a standard deviation of
1.7 degrees for [NII] and 2.4 degrees for \halp.
The inclinations were derived after excluding all inner measurements
($R<8$\arcsec; likely affected by the warp), and after an iterative three-sigma clipping
to remove outliers.
Combining the results from the two tracers, the derived gas-disk inclination (mean and
error in mean) is $i_{\rm kin}$=13.7 $\pm$ 0.4 degrees.
The derived inclination agrees well with the value of 14\arcdeg\ found by \cite{Munoz2015}
using Spitzer photometry, and also with $i_{\rm TF}=15\arcdeg \pm 1\arcdeg$ found from
inverting the Tully-Fisher relation of \cite{verheyen2001b}
\citep[calculated as in][]{martinsson2013a}.

%

\begin{figure}
\centering
\includegraphics[width=0.50\textwidth]{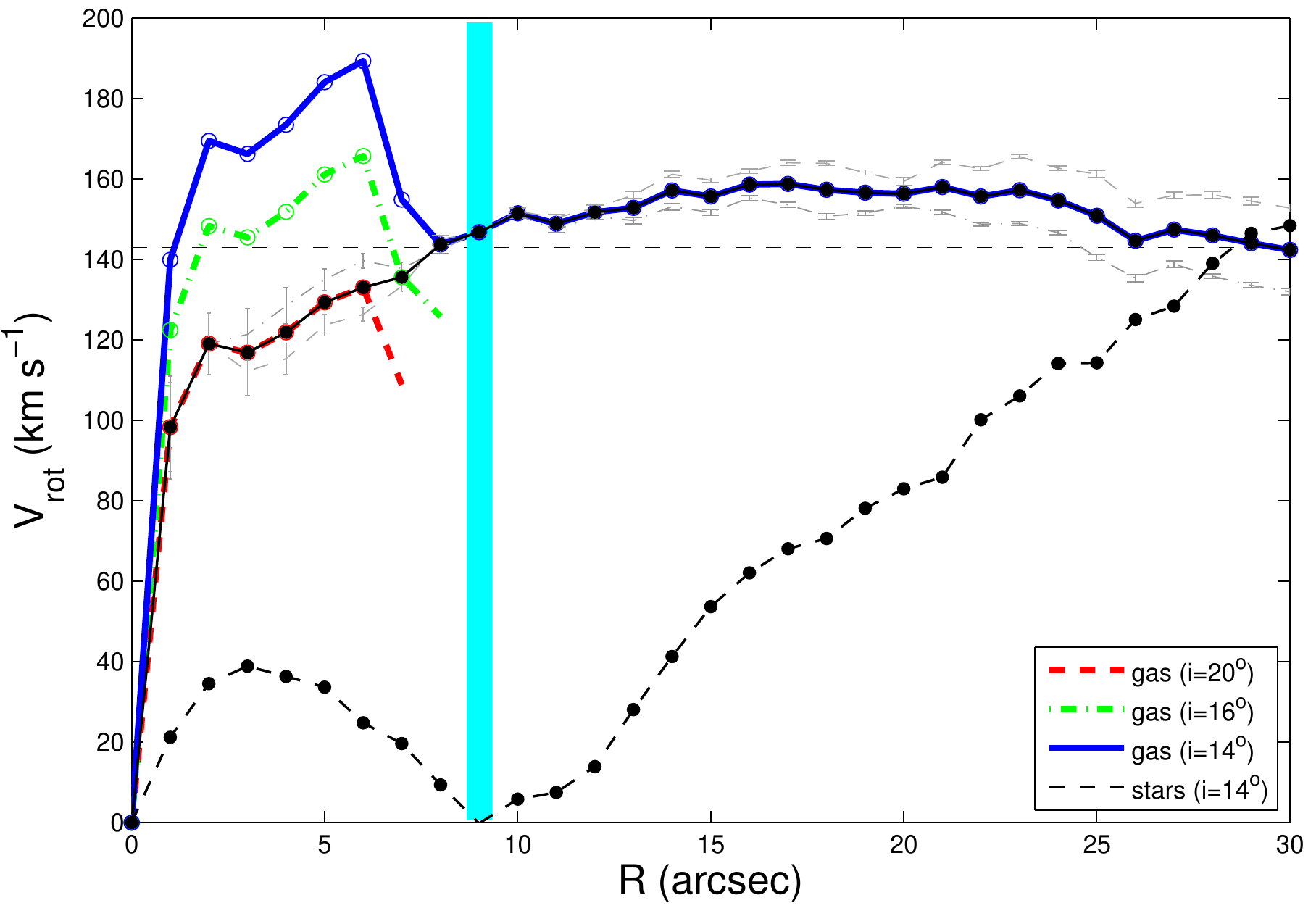}
\caption{Gas and stellar rotation curves.
The coloured lines show fitted gas (combined [NII] and \halp) rotation curves assuming
three different disk inclinations (solid blue: $i$=14\arcdeg;
dash-dotted green: $i$=16\arcdeg; dashed red: $i$=20\arcdeg).
The two thin grey lines with errorbars indicate the individual [NII] and \halp\ rotation
curves.
The dashed black line shows the fitted stellar rotation curve ($i$=14\arcdeg).
The thin dashed black horizontal line indicates the rotation speed expected from the
inverse Tully-Fisher relation. The broad vertical line indicates the ring radius
\citep[$R$$=$$9\farcs0$; ][]{comeron2010}.
}
\label{fig:RCs}
\end{figure}

Considering a possible inner warp, in Fig.~\ref{fig:RCs} we plot the derived rotation
curve with a constant inclination of 14\arcdeg\ as a blue solid line.
The derived rotation curve indeed shows a significant peak around R=2--6\arcsec. However,
by setting the inclination of the tilted rings in the inner 6\arcsec\ to 20\arcdeg, and
16\arcdeg\ in the intermediate region at 7\arcsec, we obtain a rotation curve that is
continuously rising; more what one would expect for a disk galaxy with a small inner mass
concentration.

\cite{Silchenko_Moiseev2006} found that the innermost ($R<3$\arcsec) gas disk is
strongly warped, with an inner disk inclination $i_{\rm gas} > 35$\arcdeg.
This could be a more strongly inclined circumnuclear disk
(see also Sect.~\ref{sec:GasKin}), separated from, but possibly affecting, the outer disk
for which we find has only a modest warp of 6\arcdeg.

%
%

%
\subsection{Stellar kinematic decomposition of the ring}
\label{sec:RingDec}
In Sect.~\ref{sec:StarKin}, we noted the very low stellar velocities in the ring region,
in particular along the kinematic major axis (Fig.~\ref{fig:star_kin}).
The stellar rotation curve (Fig.~\ref{fig:RCs}), derived in the same way as the gas
rotation curve\footnote{We find a stellar disk inclination of 12.1$\pm$4.9
degrees, but here adopt $i$=14\arcdeg. The derived PA of the outer (R$>$14\arcsec)
stellar disk is 321$\pm$1 degrees, i.e. $\sim$180\arcdeg\ difference from the gas disk,
as expected for aligned but counter-rotating disks.}
(Sect.~\ref{sec:warp}), indeed behaves oddly with
a broad dip at the radius of the ring.
This is because thus far we have fitted the stellar kinematics assuming a single
line-of-sight velocity distribution (LOSVD), whereas what we observe is the superposition
of two distinct LOSVDs, one of the rotating old stars in the bulge and disk and the
other of the counter-rotating young stars that formed mainly in the ring. When
interpreting such a superposition as a single velocity distribution common to all stars
along the line of sight, this leads to an LOSVD estimate centered near the average
velocity from the two old and young stellar populations, which is close to zero in this
case since these are counter-rotating, and also results in an increase in the
measured velocity dispersion due to the shifts between the lines.

One can separate the rotation of the old component from that of the newly-formed stars in
the ring by fitting two LOSVDs in the pPXF code, as done in \cite{coccato2011} and
more recently in \cite{morelli2017}.
For this we used our 12 high-S/N ring apertures and the result of our stellar-population
analysis (see Sect.~\ref{sec:StarPops}) in order to create a ``bulge'' and a ``ring''
template by combining all single-age templates older and younger than 1~Gyr,
respectively, according to their best fitting mass weights (see Fig.~\ref{Fig:SFH_a}).
Due to the low inclination of NGC~7742, and the consequent small velocity
separation between the old and young stellar component, we found it necessary to
impose to the young stars in the ring the same velocity as observed for the ionised
(\halp) gas in order to converge to sensible results.

From the two-disk profile of \cite{Silchenko_Moiseev2006} indicated in our
Fig.~\ref{Fig:SNa}, we estimate that the disk light contributes $\sim$67\% of the R-band
light in the ring region (7\arcsec{--}12\arcsec), with the remaining 33\% most likely
belonging to recently formed stars contributing to the observed surface-brightness bump
(the bulge light is not contributing much at this radius). This is consistent with both
the result from the kinematic decomposition in this section and our stellar-population
analysis in Sec.~\ref{sec:StarPops}, where stars younger than 1~Gyr account to $\sim$35\%
and $\sim$32\% of the light in the R-band, respectively.

\begin{figure}
\centering
\includegraphics[width=0.5\textwidth]{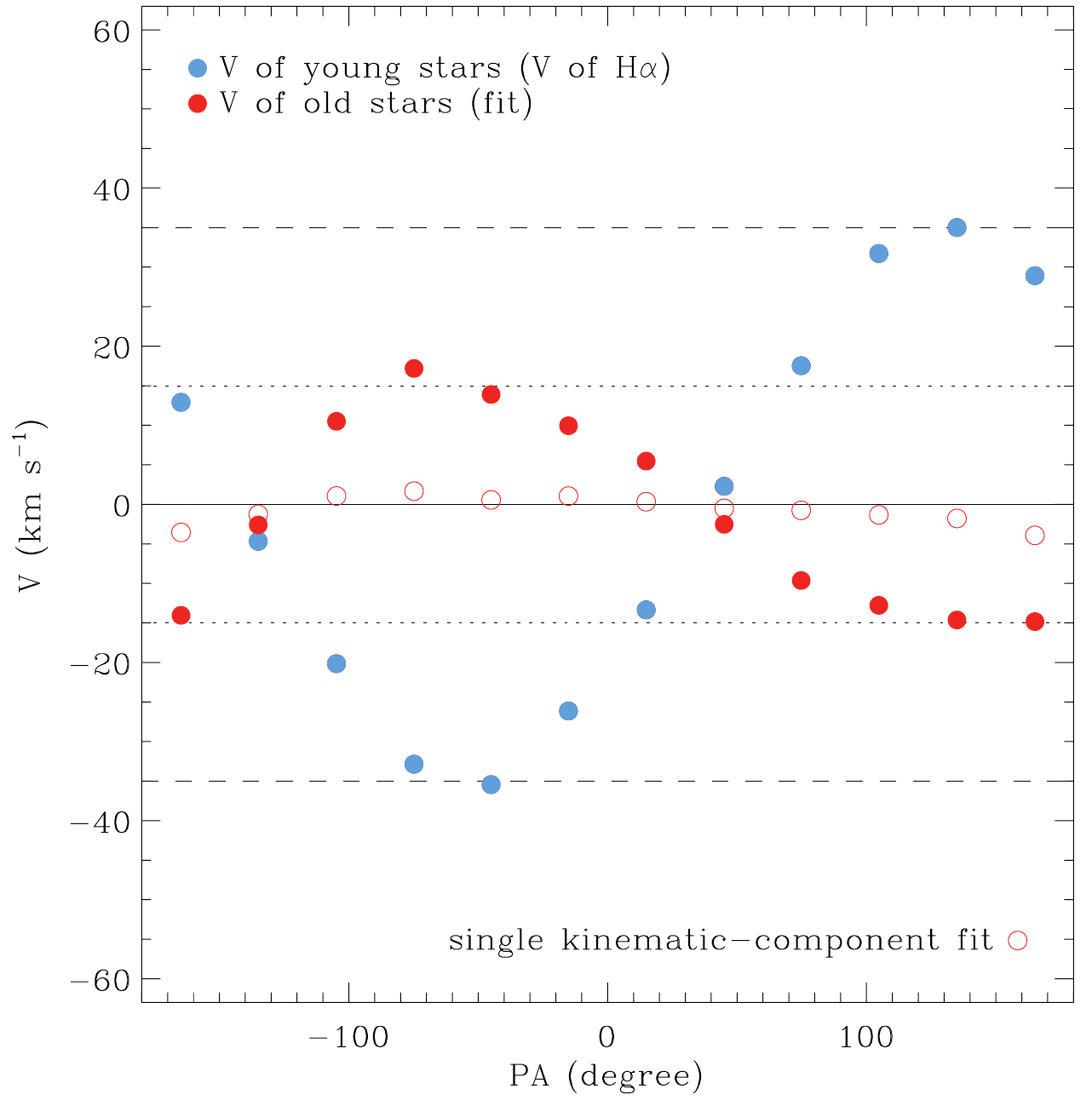}
\caption{Two-component fit to the stellar kinematics, plotted azimuthally along the ring.
Empty red circles show the line-of-sight rotational velocity derived from a single
kinematic component fit, while the filled red and blue circles show the rotation derived
from the old stars and young stars (using \halp\ as a proxy), respectively.
}
\label{fig:star_kin_2comp}
\end{figure}

Figure~\ref{fig:star_kin_2comp} shows the recovered velocity of the two components
azimuthally along the ring, as well as the velocity derived from the single-component fit
(which was used to derive the stellar rotation curve in Fig.~\ref{fig:RCs}). 
It is evident that an old stellar component and a young stellar component,
counter-rotating in the same way as the ionized gas, can reproduce the observed
single-component fit of nearly zero amplitude.

The superposition of the old and counter-rotating young components would also
cause an overall broader line-of-sight velocity dispersion, in particular along the line
of the nodes where co- and counter-rotating velocities would differ most. That is how the
enhancement feature in the velocity dispersion maps can be
interpreted, which is indeed aligned with the direction of the kinematics major axis in
both the stellar and gas maps.

%
%

%
\subsection{Stellar populations and star-formation histories}
\label{sec:StarPops}
As described in Sect.~\ref{sec:Analysis}, prior to extracting the stellar kinematics and 
carrying out our emission-line measurements in the Voronoi-binned spectra, we used pPXF
and GandALF and the stellar-population model library of \cite{Falcon-Barroso2011} to fit
29 high-S/N spectra extracted within the apertures shown in Fig.~\ref{fig:psf_Apertures}.
While this allows to use a more restricted set of optimal templates during the pPXF and
GandALF fit to our Voronoi bins, thus speeding up this process, the results of our
aperture fits also allow us to study the star-formation
history (SFH) in the nuclear, bulge, ring, and disk regions
(but see Sect.~\ref{sec:Analysis} on the definition of these regions).
The SFH can indeed be reconstructed using the weights assigned to each of the MILES
single-age stellar-population models while fitting the aperture spectra, and more
specifically when using GandALF, since during such a fit we use only multiplicative
polynomials.

%
\begin{figure}[t]
\centering
\includegraphics[width=0.5\textwidth]{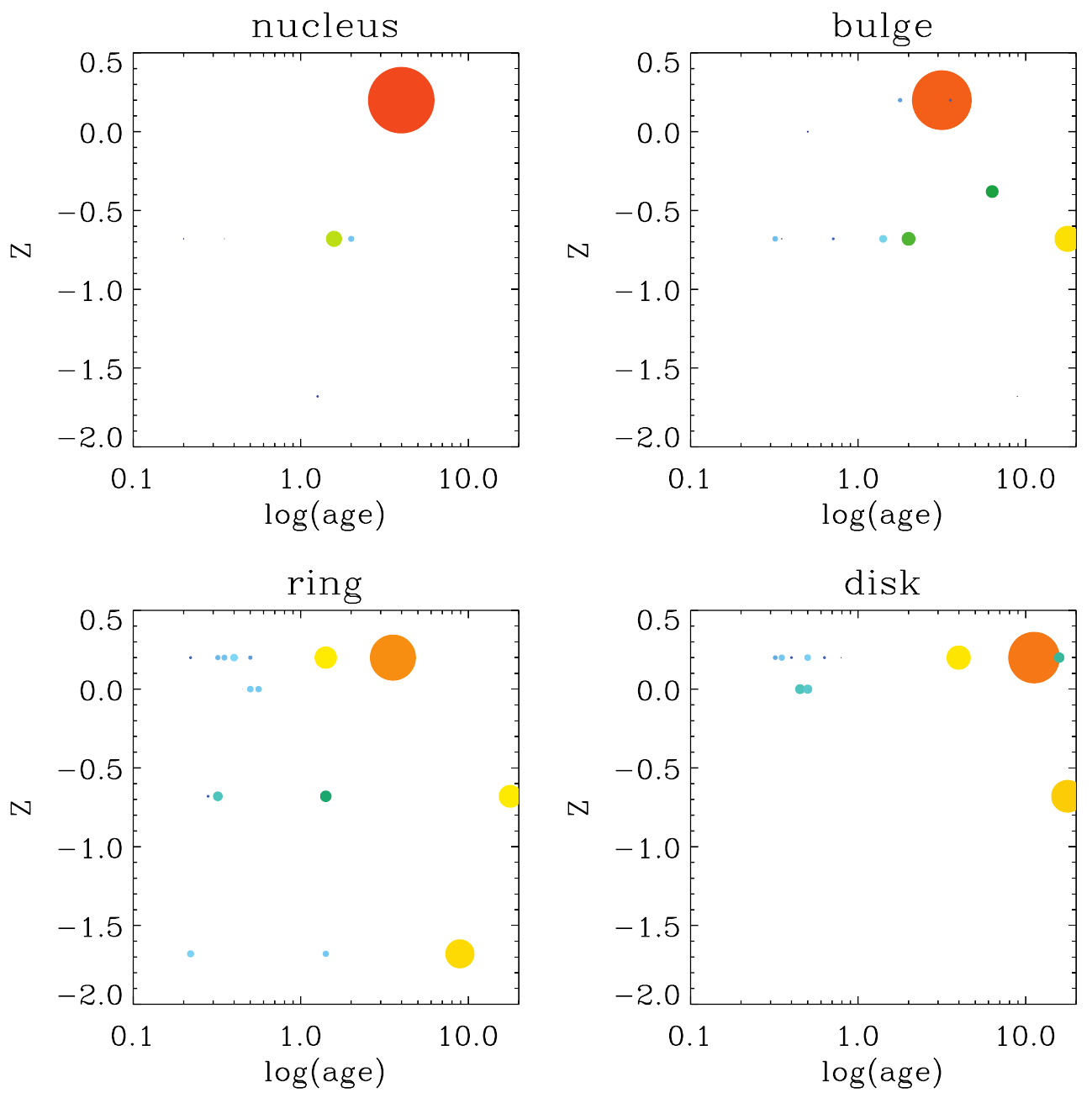}
\caption{Stellar-population properties (age and metallicity) in the nuclear, bulge, ring,
and disk regions. Each dot shows the mass-weight assigned to the model used during the
GandALF fit for the various apertures, where a larger and redder dot indicates a larger
mass-weight.
}
\label{Fig:SFH_a}
\end{figure}
%
%
%
\begin{figure}[t]
\centering
\includegraphics[width=0.50\textwidth]{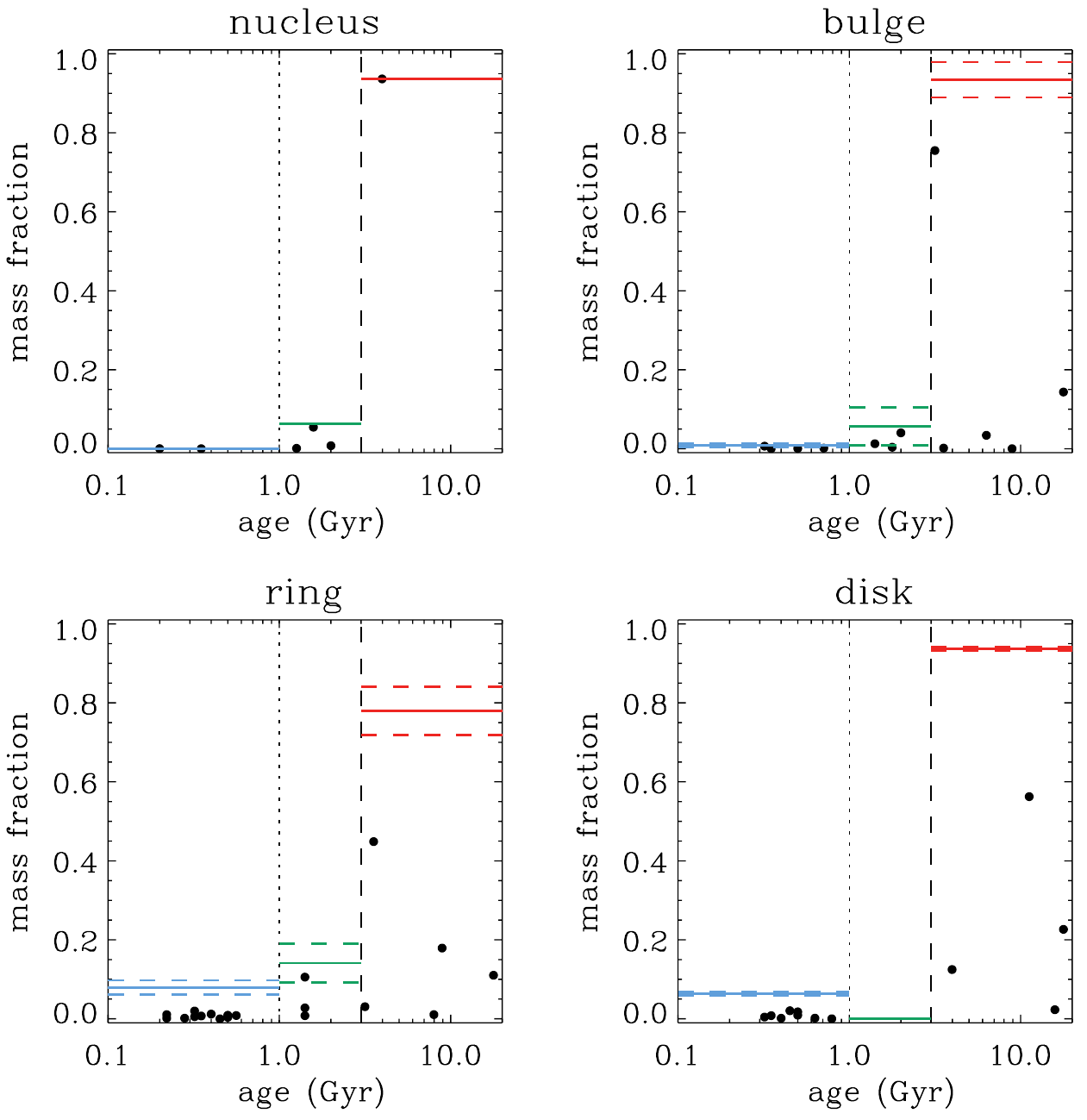}
\caption{Star-formation histories of the nucleus, bulge, ring, and disk regions.
The black dots correspond to the the coloured dots in Fig.~\ref{Fig:SFH_a}. The locations
of the horizontal blue, green, and red lines indicate the mass fraction of the young
($<$1~Gyr), intermediate-old (1--3~Gyr), and old ($>$3~Gyr) stellar populations,
respectively, calculated by adding the the contributions from individual templates (black
dots).
The horizontal dashed lines indicate the formal errors in the mass fractions, deduce
by treating each different azimuthal aperture spectrum as an independent probe of the
bulge, ring, and disk regions. For the nucleus, errors should be taken to be similar to
those reported for the bulge regions, which are similarly dominated by old stellar
populations. Note that the disk measurements are very robust, whereas stochastic effects
such as the presence of different numbers of HII regions contribute to the larger errors
in the stellar-population analysis for the ring region.
}
\label{Fig:SFH_b}
\end{figure}

Fig.~\ref{Fig:SFH_a} shows, as a function of their age, the mass-weight assigned to each
of the MILES models that were used during the GandALF fit for the nuclear, bulge, ring,
and disk apertures, without distinguishing the results from individual apertures covering
the bulge, ring, and disk regions (that is, 4, 12, and 12, respectively). 
This was done on purpose, since we do not expect to be able to ascertain the presence of
possible azimutal gradients (for instance, in the ring regions the circular orbital period
is just $\sim$20~Myr, well below the age of the youngest of our single-age templates).
On the other hand, by effectively considering our different aperture spectra for the
bulge, ring, and disk regions as independent probes of these regions we can instead convey
in a conservative way the typical uncertainties in the template-weighting process that
enters into the GandALF procedures.

To ease the reading of the SFH in NGC~7742, in Fig.~\ref{Fig:SFH_b} we group together the
contribution of young (up to 1 Gyr old), intermediate (between 1 and 3 Gyr old) and old
(above 3 Gyr) stellar populations. In this way, we see that young stars are predominantly
found both in the ring and disk regions of NGC~7742, where they contribute, respectively,
8\% and 6\% to the total stellar mass, although some young stars (up to 1\%) also appear
in the bulge. Intermediate-age populations are also present in the ring regions, where
they add up to 14\% of the mass, as well as in the bulge and nuclear
regions (although to a lower extent). 
This is consistent with previous studies advocating a prolonged SFH in nuclear rings
\citep[e.g.,][]{allard2006, sarzi2007}, and further suggests that stars formed in the ring
may also migrate out of these structures. Interestingly, there is little or no evidence
for intermediate-age stars in the disk regions, beyond the central ring. We speculate that
this is an indication that whereas star formation in the ring was triggered soon after
the merging event that NGC~7742 may have experienced 2--3~Gyr ago, gas in the outer
regions took a longer time to settle down so that in the disk star formation only started
in the past 1~Gyr.
%

%
\section{Summary}
\label{sec:Conclusions}
We have presented results from MUSE observations of the non-barred, nuclear-ring galaxy
NGC~7742.
The large FoV of MUSE allowed us to map the gas kinematics much further out
in the disk than what has been possible with previous data sets, while the high spatial
resolution made it possible to derive the kinematics in the central parts of the galaxy.

From our tilted-ring analysis, we derived a gas-disk inclination of
$i_{\rm kin}$=13.7 $\pm$ 0.4 degrees beyond the radius $R\approx8$\arcsec.
Within this radius, we identified an inner warp of the gas disk in the order of
6\arcdeg, which due to projection causes a peak in the ionized-gas rotation curve around
$R=6\arcsec$.

We confirmed the counter-rotation of gas and stars which presumably has a minor
merger origin.
The recently-formed young stars, which are concentrated in (but not limited to) the ring,
co-rotate with the gas and thus counter-rotate with respect to the old stars. A simple
one-component fit to the stellar kinematics erroneously yields a low-amplitude
stellar rotation curve, as highlighted when we combine the kinematics of the old stellar
component with the counter-rotating young component assumed to move as the ionized gas
to reproduce the observed one-component velocity in the ring.

We studied the SFH, using high-S/N spectra extracted within large apertures in the
nuclear, bulge, ring, and disk regions. Our results suggest a prolonged SFH in the nuclear
ring over more than 1~Gyr.
However, the lack of intermediate-age (1--3~Gyr) stars in the disk regions indicates that,
although star formation in the ring was presumably triggered soon after the merging event
that NGC~7742 experienced 2--3 Gyr ago, the gas in the outer regions took longer time
to settle down and the star formation in the disk therefore started much later.
%

%
\begin{acknowledgements}
We want to thank the anonymous referee for the constructive comments and suggestions
that helped improving this paper.
We also thank Nate Bastian at ARI of Liverpool John Moores University for discussions and
comments on the manuscript.
T.P.K.M., J.H.K., and J.F-B.\ acknowledge financial support from the Spanish Ministry of
Economy and Competitiveness (MINECO) under grant numbers AYA2013-41243-P, AYA2016-76219-P,
and AYA2016-77237-C3-1-P.
T.P.K.M.\ is also supported by grant ESP2015-68964 from the Ministerio de Econom\'{i}a y
Competitividad of Spain.
J.H.K.\ acknowledges financial support from the European Union’s Horizon 2020 research and
innovation programme under the Marie Sk{\l}odowska-Curie grant agreement No~721463 to the
SUNDIAL ITN network.
This research has made use of the NASA/IPAC Extragalactic Database (NED) which is
operated by the Jet Propulsion Laboratory, California Institute of Technology, under
contract with the National Aeronautics and Space Administration.
Based on observations made with the NASA/ESA Hubble Space Telescope, and obtained from the
Hubble Legacy Archive, which is a collaboration between the Space Telescope Science
Institute (STScI/NASA), the European Space Agency (ST-ECF/ESAC/ESA) and the Canadian
Astronomy Data Centre (CADC/NRC/CSA).
Based on observations collected at the European Organisation for Astronomical Research in
the Southern Hemisphere under ESO programme 60.A-9301(A).
\end{acknowledgements}
%

\bibliography{./Martinsson_N7742}

\begin{thebibliography}{41}
\expandafter\ifx\csname natexlab\endcsname\relax\def\natexlab#1{#1}\fi

\bibitem[{{Allard} {et~al.}(2006){Allard}, {Knapen}, {Peletier}, \&
  {Sarzi}}]{allard2006}
{Allard}, E.~L., {Knapen}, J.~H., {Peletier}, R.~F., \& {Sarzi}, M. 2006,
  \mnras, 371, 1087

\bibitem[{{Andersen} \& {Bershady}(2013)}]{andersen2013}
{Andersen}, D.~R. \& {Bershady}, M.~A. 2013, \apj, 768, 41

\bibitem[{{Bacon} {et~al.}(2010){Bacon}, {Accardo}, {Adjali}, {Anwand},
  {Bauer}, {Biswas}, {Blaizot}, {Boudon}, {Brau-Nogue}, {Brinchmann},
  {Caillier}, {Capoani}, {Carollo}, {Contini}, {Couderc}, {Daguis{\'e}},
  {Deiries}, {Delabre}, {Dreizler}, {Dubois}, {Dupieux}, {Dupuy}, {Emsellem},
  {Fechner}, {Fleischmann}, {Fran{\c c}ois}, {Gallou}, {Gharsa}, {Glindemann},
  {Gojak}, {Guiderdoni}, {Hansali}, {Hahn}, {Jarno}, {Kelz}, {Koehler},
  {Kosmalski}, {Laurent}, {Le Floch}, {Lilly}, {Lizon}, {Loupias}, {Manescau},
  {Monstein}, {Nicklas}, {Olaya}, {Pares}, {Pasquini}, {P{\'e}contal-Rousset},
  {Pell{\'o}}, {Petit}, {Popow}, {Reiss}, {Remillieux}, {Renault}, {Roth},
  {Rupprecht}, {Serre}, {Schaye}, {Soucail}, {Steinmetz}, {Streicher}, {Stuik},
  {Valentin}, {Vernet}, {Weilbacher}, {Wisotzki}, \& {Yerle}}]{bacon2010}
{Bacon}, R., {Accardo}, M., {Adjali}, L., {et~al.} 2010, in Society of
  Photo-Optical Instrumentation Engineers (SPIE) Conference Series, Vol. 7735,
  Society of Photo-Optical Instrumentation Engineers (SPIE) Conference Series

\bibitem[{{Bassett} {et~al.}(2017){Bassett}, {Bekki}, {Cortese}, \&
  {Couch}}]{bassett2017}
{Bassett}, R., {Bekki}, K., {Cortese}, L., \& {Couch}, W. 2017, \mnras, 471,
  1892

\bibitem[{{Begeman}(1989)}]{begeman1989}
{Begeman}, K.~G. 1989, \aap, 223, 47

\bibitem[{{Bekki}(1998)}]{bekki1998}
{Bekki}, K. 1998, \apjl, 502, L133

\bibitem[{{Bertola} {et~al.}(1996){Bertola}, {Cinzano}, {Corsini}, {Pizzella},
  {Persic}, \& {Salucci}}]{bertola1996}
{Bertola}, F., {Cinzano}, P., {Corsini}, E.~M., {et~al.} 1996, \apjl, 458, L67

\bibitem[{{Buta} \& {Combes}(1996)}]{buta1996}
{Buta}, R. \& {Combes}, F. 1996, \fcp, 17, 95

\bibitem[{{Cappellari} \& {Copin}(2003)}]{cappellari2003}
{Cappellari}, M. \& {Copin}, Y. 2003, \mnras, 342, 345

\bibitem[{{Cappellari} \& {Emsellem}(2004)}]{cappellari2004}
{Cappellari}, M. \& {Emsellem}, E. 2004, \pasp, 116, 138

\bibitem[{{Coccato} {et~al.}(2011){Coccato}, {Morelli}, {Corsini}, {Buson},
  {Pizzella}, {Vergani}, \& {Bertola}}]{coccato2011}
{Coccato}, L., {Morelli}, L., {Corsini}, E.~M., {et~al.} 2011, \mnras, 412,
  L113

\bibitem[{{Combes} \& {Gerin}(1985)}]{combes1985}
{Combes}, F. \& {Gerin}, M. 1985, \aap, 150, 327

\bibitem[{{Comer{\'o}n} {et~al.}(2010){Comer{\'o}n}, {Knapen}, {Beckman},
  {Laurikainen}, {Salo}, {Mart{\'{\i}}nez-Valpuesta}, \& {Buta}}]{comeron2010}
{Comer{\'o}n}, S., {Knapen}, J.~H., {Beckman}, J.~E., {et~al.} 2010, \mnras,
  402, 2462

\bibitem[{{Corsini}(2014)}]{corsini2014}
{Corsini}, E.~M. 2014, in Astronomical Society of the Pacific Conference
  Series, Vol. 486, Multi-Spin Galaxies, ed. E.~{Iodice} \& E.~M. {Corsini}, 51

\bibitem[{{Corsini} \& {Bertola}(1998)}]{CorsiniBertola1998}
{Corsini}, E.~M. \& {Bertola}, F. 1998, Journal of Korean Physical Society, 33,
  S574

\bibitem[{{de Zeeuw} {et~al.}(2002){de Zeeuw}, {Bureau}, {Emsellem}, {Bacon},
  {Carollo}, {Copin}, {Davies}, {Kuntschner}, {Miller}, {Monnet}, {Peletier},
  \& {Verolme}}]{deZeeuw2002}
{de Zeeuw}, P.~T., {Bureau}, M., {Emsellem}, E., {et~al.} 2002, \mnras, 329,
  513

\bibitem[{{Falc{\'o}n-Barroso} {et~al.}(2006){Falc{\'o}n-Barroso}, {Bacon},
  {Bureau}, {Cappellari}, {Davies}, {de Zeeuw}, {Emsellem}, {Fathi},
  {Krajnovi{\'c}}, {Kuntschner}, {McDermid}, {Peletier}, \&
  {Sarzi}}]{Falcon-Barroso2006}
{Falc{\'o}n-Barroso}, J., {Bacon}, R., {Bureau}, M., {et~al.} 2006, \mnras,
  369, 529

\bibitem[{{Falc{\'o}n-Barroso} {et~al.}(2011){Falc{\'o}n-Barroso},
  {S{\'a}nchez-Bl{\'a}zquez}, {Vazdekis}, {Ricciardelli}, {Cardiel}, {Cenarro},
  {Gorgas}, \& {Peletier}}]{Falcon-Barroso2011}
{Falc{\'o}n-Barroso}, J., {S{\'a}nchez-Bl{\'a}zquez}, P., {Vazdekis}, A.,
  {et~al.} 2011, \aap, 532, A95

\bibitem[{{Fogarty} {et~al.}(2015){Fogarty}, {Scott}, {Owers}, {Croom},
  {Bekki}, {Houghton}, {van de Sande}, {D'Eugenio}, {Cecil}, {Colless},
  {Bland-Hawthorn}, {Brough}, {Cortese}, {Davies}, {Jones}, {Pracy}, {Allen},
  {Bryant}, {Goodwin}, {Green}, {Konstantopoulos}, {Lawrence}, {Lorente},
  {Richards}, \& {Sharp}}]{fogarty2015}
{Fogarty}, L.~M.~R., {Scott}, N., {Owers}, M.~S., {et~al.} 2015, \mnras, 454,
  2050

\bibitem[{{Freudling} {et~al.}(2013){Freudling}, {Romaniello}, {Bramich},
  {Ballester}, {Forchi}, {Garc{\'{\i}}a-Dabl{\'o}}, {Moehler}, \&
  {Neeser}}]{freudling2013}
{Freudling}, W., {Romaniello}, M., {Bramich}, D.~M., {et~al.} 2013, \aap, 559,
  A96

\bibitem[{{Heller} \& {Shlosman}(1994)}]{heller1994}
{Heller}, C.~H. \& {Shlosman}, I. 1994, \apj, 424, 84

\bibitem[{{Hsieh} {et~al.}(2012){Hsieh}, {Ho}, {Kohno}, {Hwang}, \&
  {Matsushita}}]{hsieh2012}
{Hsieh}, P.-Y., {Ho}, P.~T.~P., {Kohno}, K., {Hwang}, C.-Y., \& {Matsushita},
  S. 2012, \apj, 747, 90

\bibitem[{{Kauffmann} {et~al.}(2003){Kauffmann}, {Heckman}, {Tremonti},
  {Brinchmann}, {Charlot}, {White}, {Ridgway}, {Brinkmann}, {Fukugita}, {Hall},
  {Ivezi{\'c}}, {Richards}, \& {Schneider}}]{kauffmann2003b}
{Kauffmann}, G., {Heckman}, T.~M., {Tremonti}, C., {et~al.} 2003, \mnras, 346,
  1055

\bibitem[{{Knapen}(2005)}]{knapen2005}
{Knapen}, J.~H. 2005, \aap, 429, 141

\bibitem[{{Knapen} {et~al.}(1995){Knapen}, {Beckman}, {Heller}, {Shlosman}, \&
  {de Jong}}]{knapen1995}
{Knapen}, J.~H., {Beckman}, J.~E., {Heller}, C.~H., {Shlosman}, I., \& {de
  Jong}, R.~S. 1995, \apj, 454, 623

\bibitem[{{Knapen} {et~al.}(2004){Knapen}, {Whyte}, {de Blok}, \& {van der
  Hulst}}]{knapen2004}
{Knapen}, J.~H., {Whyte}, L.~F., {de Blok}, W.~J.~G., \& {van der Hulst}, J.~M.
  2004, \aap, 423, 481

\bibitem[{{Kormendy} \& {Kennicutt}(2004)}]{kormendy2004}
{Kormendy}, J. \& {Kennicutt}, Jr., R.~C. 2004, \araa, 42, 603

\bibitem[{{Krajnovi{\'c}} {et~al.}(2011){Krajnovi{\'c}}, {Emsellem},
  {Cappellari}, {Alatalo}, {Blitz}, {Bois}, {Bournaud}, {Bureau}, {Davies},
  {Davis}, {de Zeeuw}, {Khochfar}, {Kuntschner}, {Lablanche}, {McDermid},
  {Morganti}, {Naab}, {Oosterloo}, {Sarzi}, {Scott}, {Serra}, {Weijmans}, \&
  {Young}}]{krajnovic2011}
{Krajnovi{\'c}}, D., {Emsellem}, E., {Cappellari}, M., {et~al.} 2011, \mnras,
  414, 2923

\bibitem[{{Martinsson} {et~al.}(2013){Martinsson}, {Verheijen}, {Westfall},
  {Bershady}, {Schechtman-Rook}, {Andersen}, \& {Swaters}}]{martinsson2013a}
{Martinsson}, T.~P.~K., {Verheijen}, M.~A.~W., {Westfall}, K.~B., {et~al.}
  2013, \aap, 557, A130

\bibitem[{{Mazzuca} {et~al.}(2006){Mazzuca}, {Sarzi}, {Knapen}, {Veilleux}, \&
  {Swaters}}]{mazzuca2006}
{Mazzuca}, L.~M., {Sarzi}, M., {Knapen}, J.~H., {Veilleux}, S., \& {Swaters},
  R. 2006, \apjl, 649, L79

\bibitem[{{Morelli} {et~al.}(2017){Morelli}, {Pizzella}, {Coccato}, {Corsini},
  {Dalla Bont{\`a}}, {Buson}, {Ivanov}, {Pagotto}, {Pompei}, \&
  {Rocco}}]{morelli2017}
{Morelli}, L., {Pizzella}, A., {Coccato}, L., {et~al.} 2017, \aap, 600, A76

\bibitem[{{Mu{\~n}oz-Mateos} {et~al.}(2015){Mu{\~n}oz-Mateos}, {Sheth},
  {Regan}, {Kim}, {Laine}, {Erroz-Ferrer}, {Gil de Paz}, {Comeron}, {Hinz},
  {Laurikainen}, {Salo}, {Athanassoula}, {Bosma}, {Bouquin}, {Schinnerer},
  {Ho}, {Zaritsky}, {Gadotti}, {Madore}, {Holwerda}, {Men{\'e}ndez-Delmestre},
  {Knapen}, {Meidt}, {Querejeta}, {Mizusawa}, {Seibert}, {Laine}, \&
  {Courtois}}]{Munoz2015}
{Mu{\~n}oz-Mateos}, J.~C., {Sheth}, K., {Regan}, M., {et~al.} 2015, \apjs, 219,
  3

\bibitem[{{Pizzella} {et~al.}(2004){Pizzella}, {Corsini}, {Vega Beltr{\'a}n},
  \& {Bertola}}]{pizzella2004}
{Pizzella}, A., {Corsini}, E.~M., {Vega Beltr{\'a}n}, J.~C., \& {Bertola}, F.
  2004, \aap, 424, 447

\bibitem[{{Sarzi} {et~al.}(2007){Sarzi}, {Allard}, {Knapen}, \&
  {Mazzuca}}]{sarzi2007}
{Sarzi}, M., {Allard}, E.~L., {Knapen}, J.~H., \& {Mazzuca}, L.~M. 2007,
  \mnras, 380, 949

\bibitem[{{Sarzi} {et~al.}(2006){Sarzi}, {Falc{\'o}n-Barroso}, {Davies},
  {Bacon}, {Bureau}, {Cappellari}, {de Zeeuw}, {Emsellem}, {Fathi},
  {Krajnovi{\'c}}, {Kuntschner}, {McDermid}, \& {Peletier}}]{sarzi2006}
{Sarzi}, M., {Falc{\'o}n-Barroso}, J., {Davies}, R.~L., {et~al.} 2006, \mnras,
  366, 1151

\bibitem[{{Sil'chenko} \& {Moiseev}(2006)}]{Silchenko_Moiseev2006}
{Sil'chenko}, O.~K. \& {Moiseev}, A.~V. 2006, \aj, 131, 1336

\bibitem[{{Sofue} {et~al.}(1999){Sofue}, {Tutui}, {Honma}, {Tomita},
  {Takamiya}, {Koda}, \& {Takeda}}]{sofue1999}
{Sofue}, Y., {Tutui}, Y., {Honma}, M., {et~al.} 1999, \apj, 523, 136

\bibitem[{{van der Hulst} {et~al.}(1992){van der Hulst}, {Terlouw}, {Begeman},
  {Zwitser}, \& {Roelfsema}}]{hulst1992}
{van der Hulst}, J.~M., {Terlouw}, J.~P., {Begeman}, K.~G., {Zwitser}, W., \&
  {Roelfsema}, P.~R. 1992, in Astronomical Society of the Pacific Conference
  Series, Vol.~25, Astronomical Data Analysis Software and Systems I, ed. D.~M.
  {Worrall}, C.~{Biemesderfer}, \& J.~{Barnes}, 131--136

\bibitem[{{Verheijen}(2001)}]{verheyen2001b}
{Verheijen}, M.~A.~W. 2001, \apj, 563, 694

\bibitem[{{Vogelaar} \& {Terlouw}(2001)}]{vogelaar2001}
{Vogelaar}, M.~G.~R. \& {Terlouw}, J.~P. 2001, in Astronomical Society of the
  Pacific Conference Series, Vol. 238, Astronomical Data Analysis Software and
  Systems X, ed. {F.~R.~Harnden Jr., F.~A.~Primini, \& H.~E.~Payne}, 358

\bibitem[{{Weilbacher}(2015)}]{weilbacher2015}
{Weilbacher}, P. 2015, in Science Operations 2015: Science Data Management - An
  ESO/ESA Workshop, held 24-27 November, 2015 at ESO Garching. Online at <A
  href=''https://www.eso.org/sci/meetings/2015/SciOps2015.html''>
  https://www.eso.org/sci/meetings/2015/SciOps2015.html</A>, id.1, 1

\end{thebibliography}
\bibliographystyle{./Martinsson_N7742}
\setlength{\bibsep}{1.3pt}

\end{document}